\documentclass[11pt,twoside            
  ]
  {article}
  
    \usepackage{asp2010}

\makeatletter
\DeclareRobustCommand\citenum
   {\begingroup
     \NAT@swatrue\let\NAT@ctype\z@\NAT@parfalse\let\textsuperscript\relax
     \NAT@citexnum[][]}
\makeatother

  \resetcounters
\def\lesssim{\mathrel{\hbox{\rlap{\hbox{\lower4pt\hbox{$\sim$}}}\hbox{$<$}}}}
\def\gtrsim{\mathrel{\hbox{\rlap{\hbox{\lower4pt\hbox{$\sim$}}}\hbox{$>$}}}}

\def\apjl{ApJL}

\def\aap{A\&A}

\begin{document}
\newcommand{\edot}{\dot{E}}
\newcommand{\pdot}{\dot{P}}
\newcommand{\lpwn}{L_{\rm pwn}}
\newcommand{\lpsr}{L_{\rm psr}}
\newcommand{\etapwn}{\eta_{\rm pwn}}
\newcommand{\etapsr}{\eta_{\rm psr}}
\newcommand{\chan}{{\sl Chandra}\/}
\newcommand{\be}{\begin{equation}}
\newcommand{\ee}{\end{equation}}
\newcommand{\xmm}{{\sl XMM-Newton\/}}
\newcommand{\suz}{{\sl Suzaku\/}}
\newcommand{\fermi}{{\sl Fermi\/}}

\title{Pulsar Wind Nebulae from X-rays to VHE $\gamma$-rays}

\author{Oleg Kargaltsev,$^1$ George G.\ Pavlov,$^{2,3}$, and Martin Durant$^1$ }
\affil{$^1$ University  of Florida, Bryant Space Center, Gainesville,
FL 32611}
\affil{$^2$ Pennsylvania State University, 525 Davey Lab., University Park,
PA 16802}
\affil{$^3$ St.-Petersburg State Polytechnic University, 195251, Russia}

\begin{abstract}
The number of plausible associations of extended VHE (TeV) sources with pulsars
 has been steadily growing, suggesting that many of these sources are 
 pulsar
wind nebulae (PWNe). Here we overview 
 the recent progress in X-ray and TeV observations of PWNe
 and summarize their properties.
\end{abstract}

Along with supernovae, pulsars produce copious amounts of relativistic particles and inject them into the Galactic medium.  The energies of  pulsar wind electrons and positrons range from $\sim$1 GeV  to   $\sim$1 PeV, placing their synchrotron and inverse Compton (IC) emission into radio-X-ray and GeV-TeV bands, respectively. This multiwavelength emission can be seen as a {\em pulsar-wind nebula} (PWN) \citep{2008AIPC..983..171K,2010AIPC.1248...25K}.  The wind particles can be trapped in the pulsar vicinity 
for $\gtrsim 10^{5}$ yr, forming relic
TeV PWNe, which appear to 
dominate the population of Galactic VHE $\gamma$-rays sources.  
 Here we provide an updated overview based on a compilation of observational properties of 85 PWNe or PWN candidates,
  71 of which have suggested associations with pulsars. Because of limited space, we restrict ourselves by graphical presentations and
short discussions of various correlations that
involve PWN luminosities and spectral slopes.

\begin{figure}[th!]
\hspace{-1.1cm}
\includegraphics[width=0.82\textwidth,angle=90]{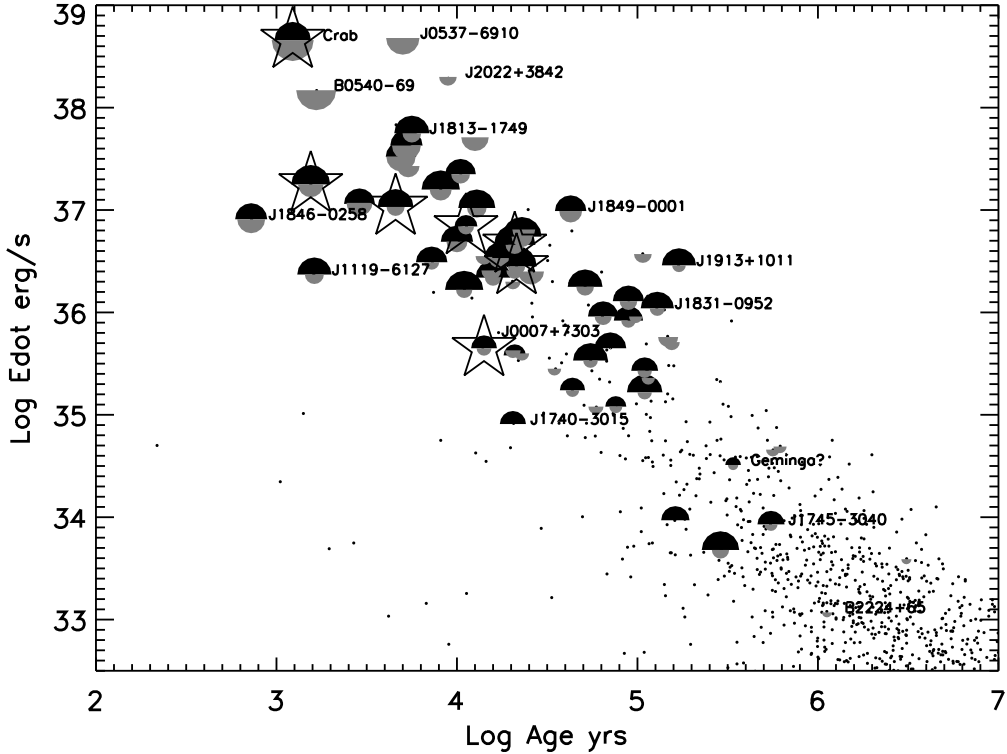}
\vspace{-0.6cm} 
\caption{ Pulsars with detected PWNe (or PWN candidates) in the $\tau$-$\edot$ diagram.
The semi-circles correspond to X-ray (grey) and TeV (black) PWNe, their sizes are proportional to logarithms of the PWN luminosities.
The small black  dots denote the pulsars from the ATNF catalog \citep{2005AJ....129.1993M}. PWNe detected by {\sl Fermi} are marked by stars.
\vspace{-0.5cm} }
\end{figure}

 \begin{figure}
 \centering
\includegraphics[width=0.7\textwidth,angle=90]{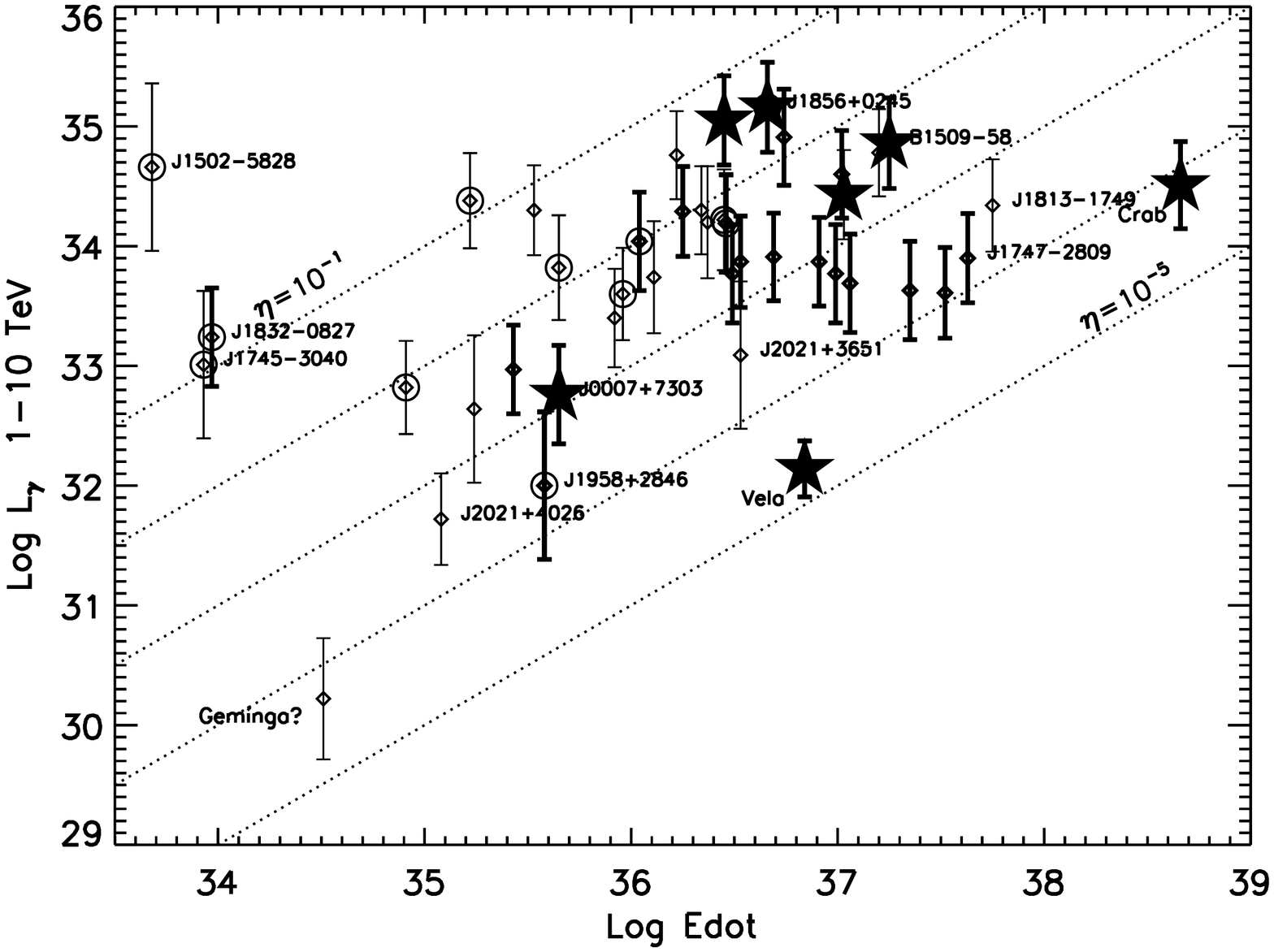}
\includegraphics[width=0.7\textwidth,angle=90]{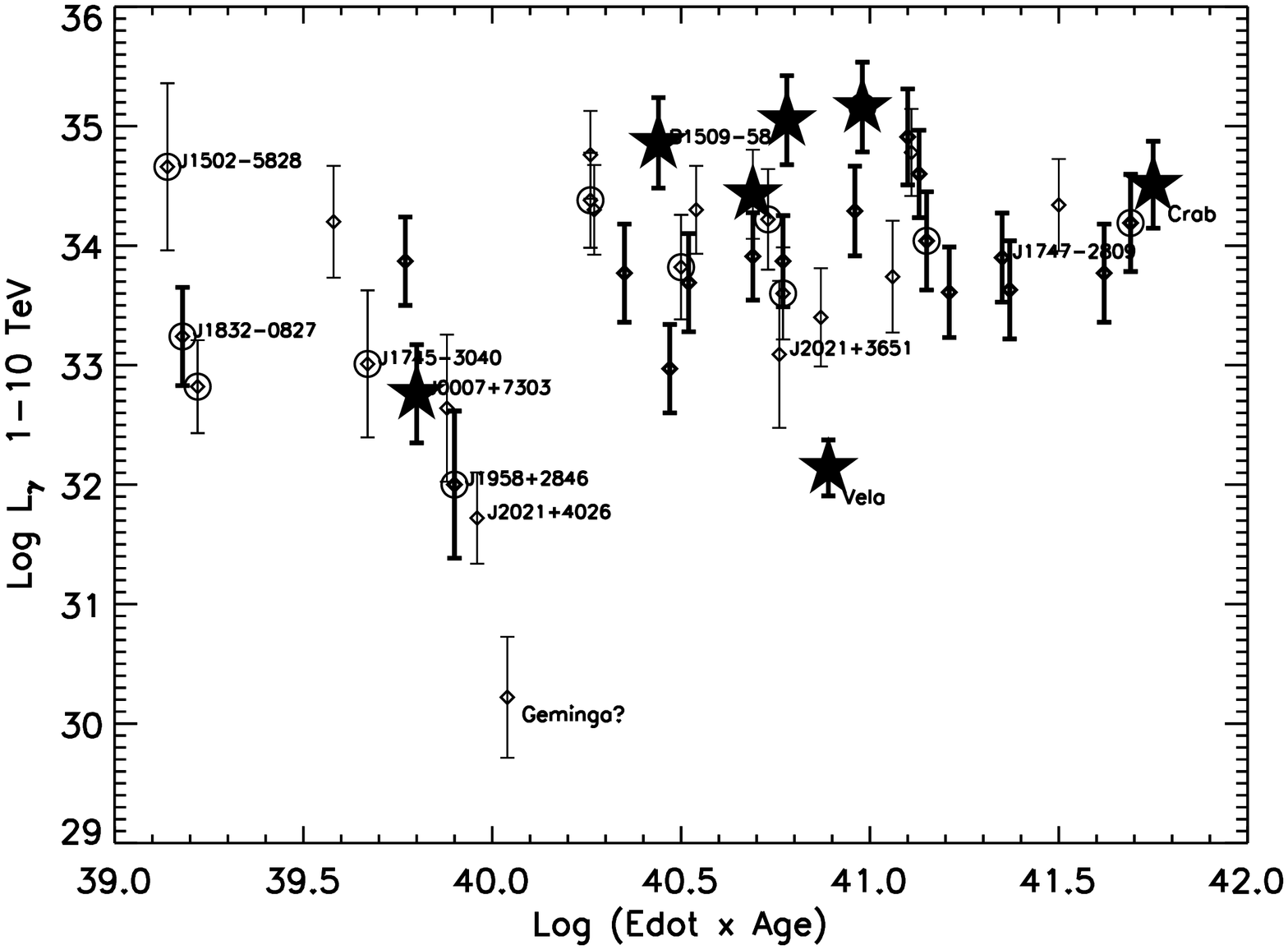}
\caption{    TeV luminosities of PWNe and PWN candidates  vs.\  pulsar's $\dot{E}$  ({\em top}) and $\edot\tau$ ({\em bottom}).  Thin error bars mark questionable associations. PWNe  undetected in X-rays are shown as circles.  PWNe detected by {\sl Fermi} are marked by stars. Dotted lines in the top panel correspond to constant 
values of the ratio $\eta_\gamma=L_\gamma/\edot$.
\vspace{-0.8cm} }
\end{figure}

In Figure 1 we show the sample of the 71  X-ray/TeV PWNe and PWN candidates, which are likely
 associated with detected pulsars, in the $\tau$-$\dot{E}$ plane,
 where $\edot$ and $\tau$ are the pulsar's spin-down power and characteristic age.
 This figure demonstrates that the X-ray/TeV PWNe are detectable up to an age of a few times $10^5$ yr, and at $\dot{E}$ as low as $\sim 10^{34}$ erg s$^{-1}$. It also demonstrates that their X-ray luminosities generally decrease with increasing age (decreasing spin-down power), while the TeV luminosities do not
show an obvious dependence on these parameters.

 \begin{figure}
 \centering
 \includegraphics[width=0.7\textwidth,angle=90]{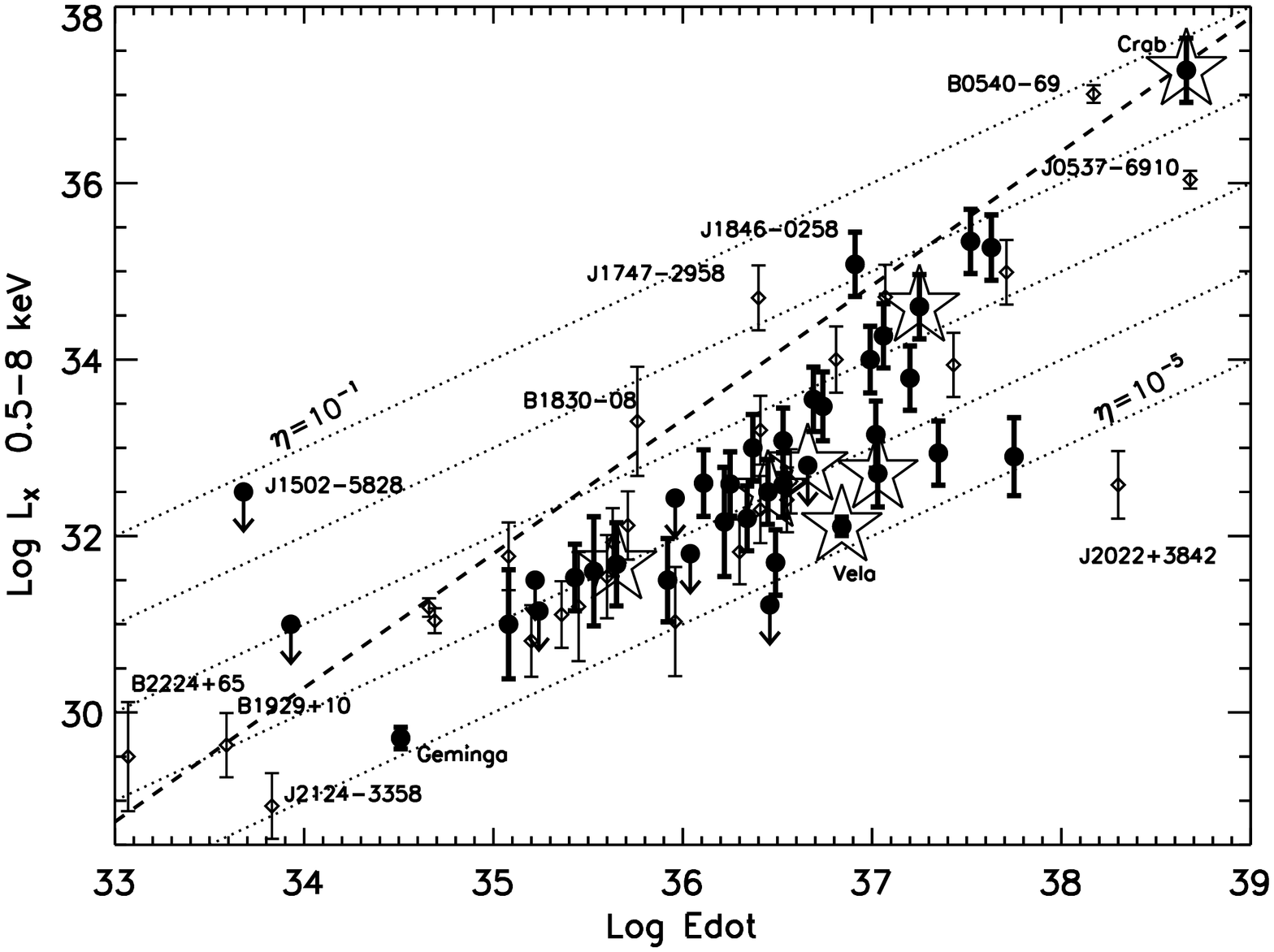}
\includegraphics[width=0.7\textwidth,angle=90]{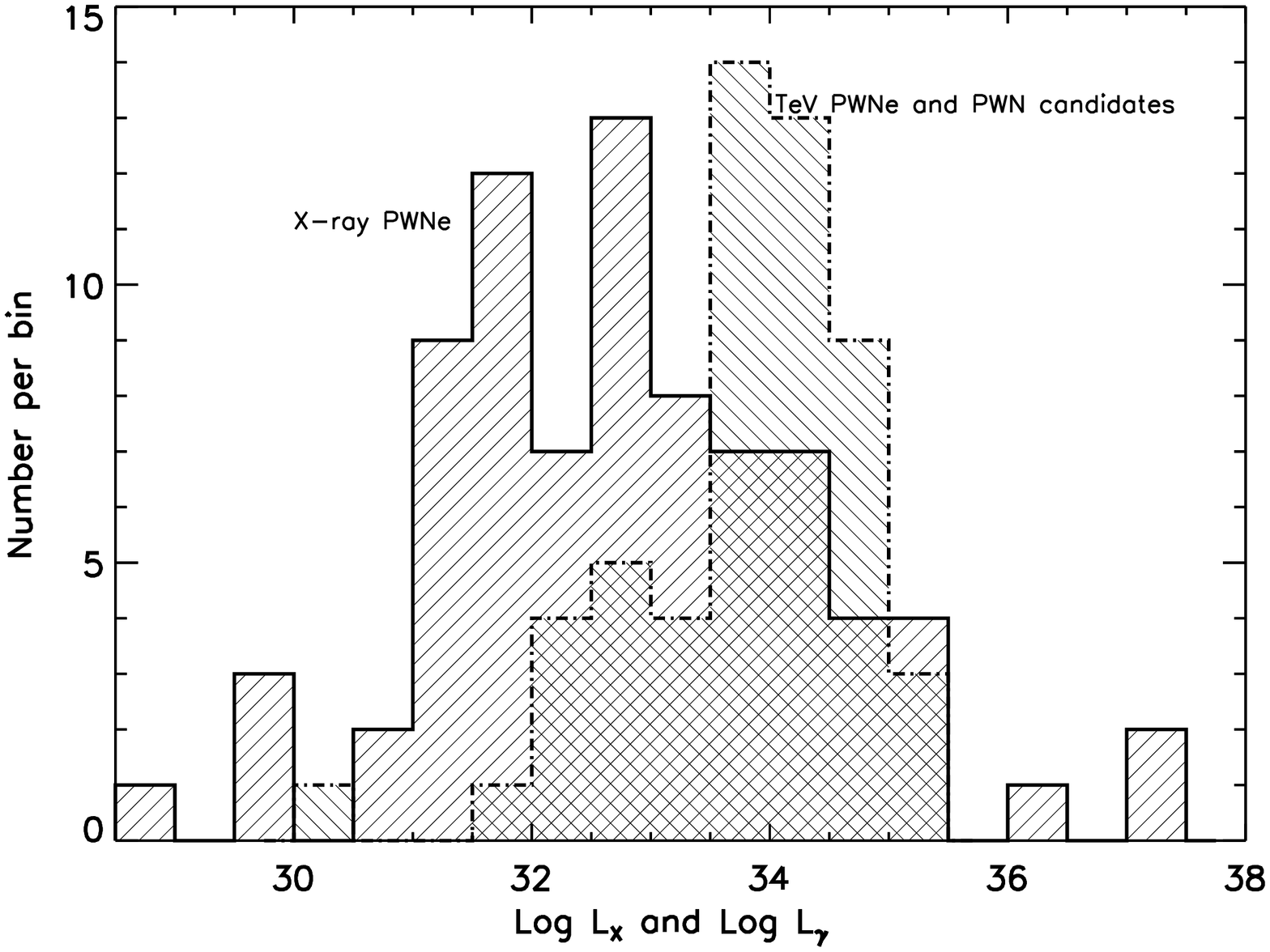}
 \caption{
{\em Top:}  X-ray luminosities of PWNe and PWN candidates
 vs.\  pulsar's $\dot{E}$.
 TeV PWNe and TeV PWN candidates  are shown as filled circles.
 Dotted straight lines correspond to constant X-ray eiffiencies; the upper bound, $\log L_X^{\rm cr} =
1.51\log\edot -21.4$ \citep{2012arXiv1202.3838K},  is shown by a dashed line. 
PWNe detected by {\sl Fermi} are marked by stars.
{\em Bottom}: Distributions of PWNe and PWNe candidates over the X-ray and TeV luminosities.
\vspace{-0.8cm} }
\end{figure}

The different correlations of the PWN TeV and X-ray 
luminosities, 
$L_\gamma$ and $L_X$, with $\dot{E}$ are 
demonstrated in top panels of Figures 2 and 3. X-ray luminosities exhibit a large spread (up to 4 orders of magnitude) 
 at a given $\dot{E}$, which apparently 
grows with
increasing $\dot{E}$. 
This means that there is no a 
unique dependence of $L_X$ on $\dot{E}$, but rather
the $L_X$ values lie below an upper bound, $L_X^{\rm cr}(\dot{E})$.
Interestingly, the bound is about the same as that found for the nonthermal 
 {\em pulsar } luminosities \citep{2012arXiv1202.3838K}.
On the contrary, 
the TeV luminosities do not show a significant correlation 
with $\dot{E}$, and the upper bound, $L_\gamma^{\rm cr} \sim 10^{35}$ erg s$^{-1}$, 
does not show a significant dependence on the pulsar's spin-down power.  
 Such 
different behavior is consistent with the interpretation of
majority of TeV PWNe as 
relic plerions, which are powered by aged leptons, injected by the pulsar long
ago, when its spin-down power was much higher. On the other hand, $L_\gamma$ does not correlate significantly with
 the product $\dot{E}\tau$ (Fig.\ 2, bottom panel), which crudely characterizes the total energy lost by the pulsar during its life time.  
The distance-independent ratio $L_\gamma/L_X$ shows a hint of growth with increasing age (decreasing $\dot{E}$) for young, powerful pulsars 
($\tau \lesssim 10$ kyr, $\dot{E}\gtrsim 10^{37}$ erg s$^{-1}$ -- see Fig.\ 4; bottom panel), 
which could, at least partly, be  
explained by the positive correlation of $L_X$ with $\edot$.
Note that most of the detected PWNe are more luminous in TeV  than in X-rays, except for
5 young objects. This trend can also be seen from the top panel of
Figure 4, which 
 includes  PWN candidates yet lacking a pulsar detection.

 \begin{figure}
 \centering
 \includegraphics[width=0.7\textwidth,angle=90]{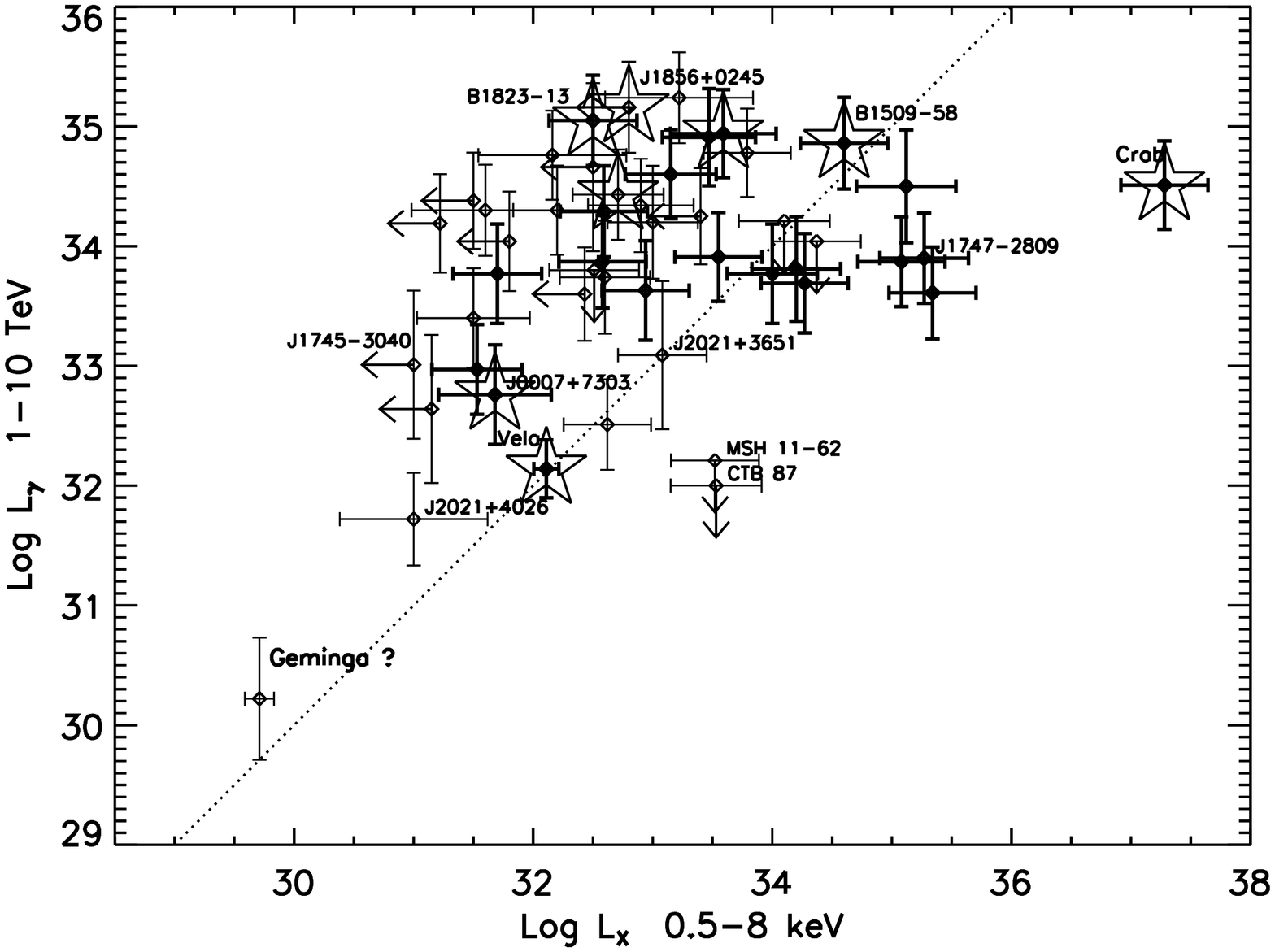}
\includegraphics[width=0.7\textwidth,angle=90]{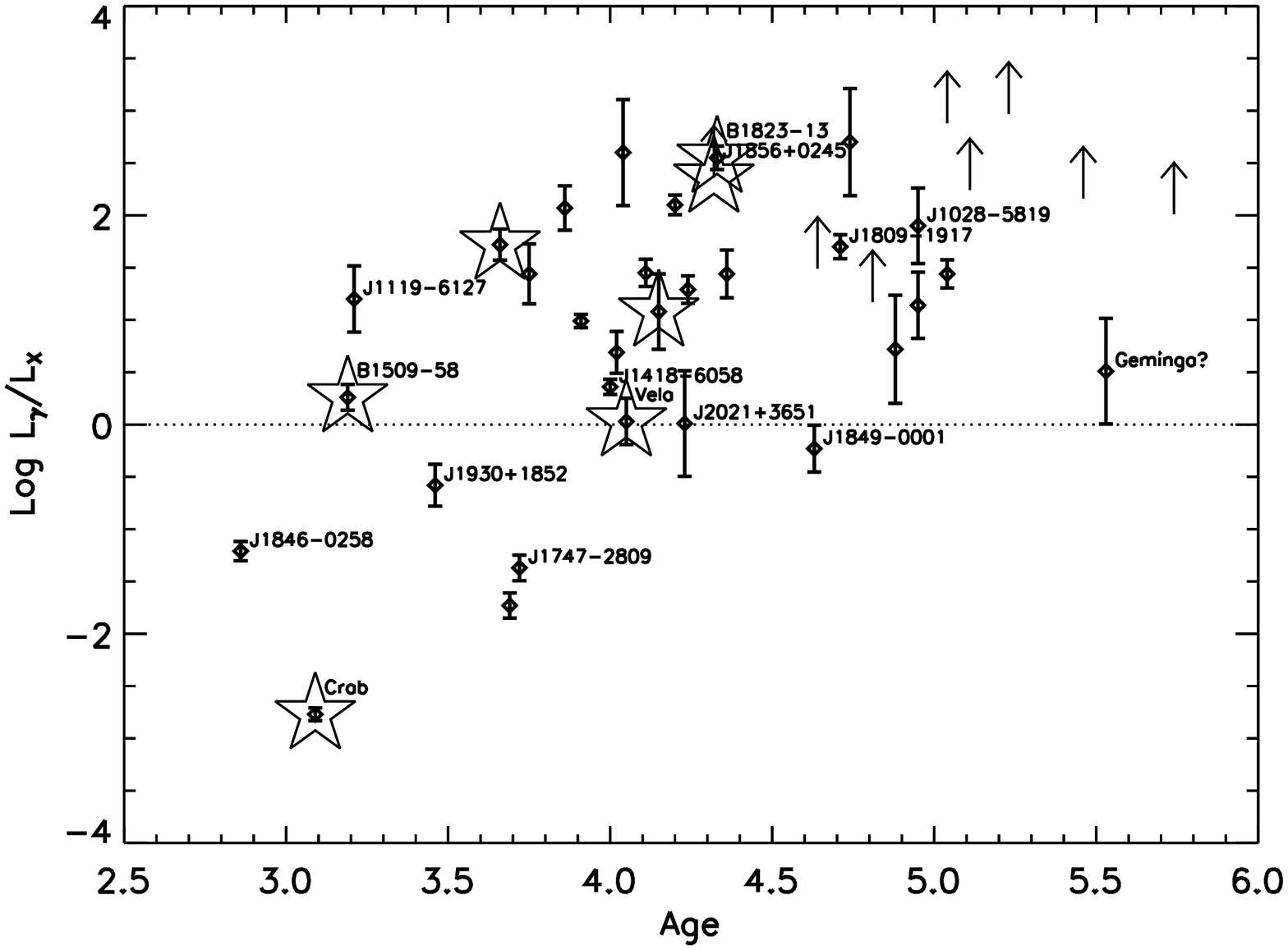}
 \caption{  
TeV luminosities vs.\ X-ray luminosities 
   ({\em top})  
 and TeV to X-ray luminosity ratios vs. pulsar's age  ({\em bottom}) 
 for PWNe and PWN candidates.  
Limits are  shown in blue.    PWNe detected by {\sl Fermi} are marked by stars.
Uncertain detctions are shown by thin lines. The dotted lines corresponds to $L_\gamma = L_X$. 
\vspace{-0.8cm} }
\end{figure}

As the spectral slopes have been measured for many PWNe in both X-ray and 
TeV bands, we can compare various correlations that involve the photon 
indices $\Gamma_X$ and $\Gamma_\gamma$. First of all, the TeV spectra are generally softer than
the X-ray spectra, with typical values $\Gamma_X \approx 1.7$ and $\Gamma_\gamma \approx 2.2$
 (Fig.\ 5; one should remember, however, that most of the TeV PWNe are significantly larger
and farther from the pulsar than their X-ray counterparts).
  Figure 5 (top panel) shows a lack of correlation between the spectral slopes of the
X-ray and TeV PWNe. We also see no correlation between $L_X$ and $\Gamma_X$
(Fig.\ 6),
while there is a hint of decreasing spread of $\Gamma_\gamma$ with increasing
$L_\gamma$.

 \begin{figure}
 \centering
 \includegraphics[width=0.7\textwidth,angle=90]{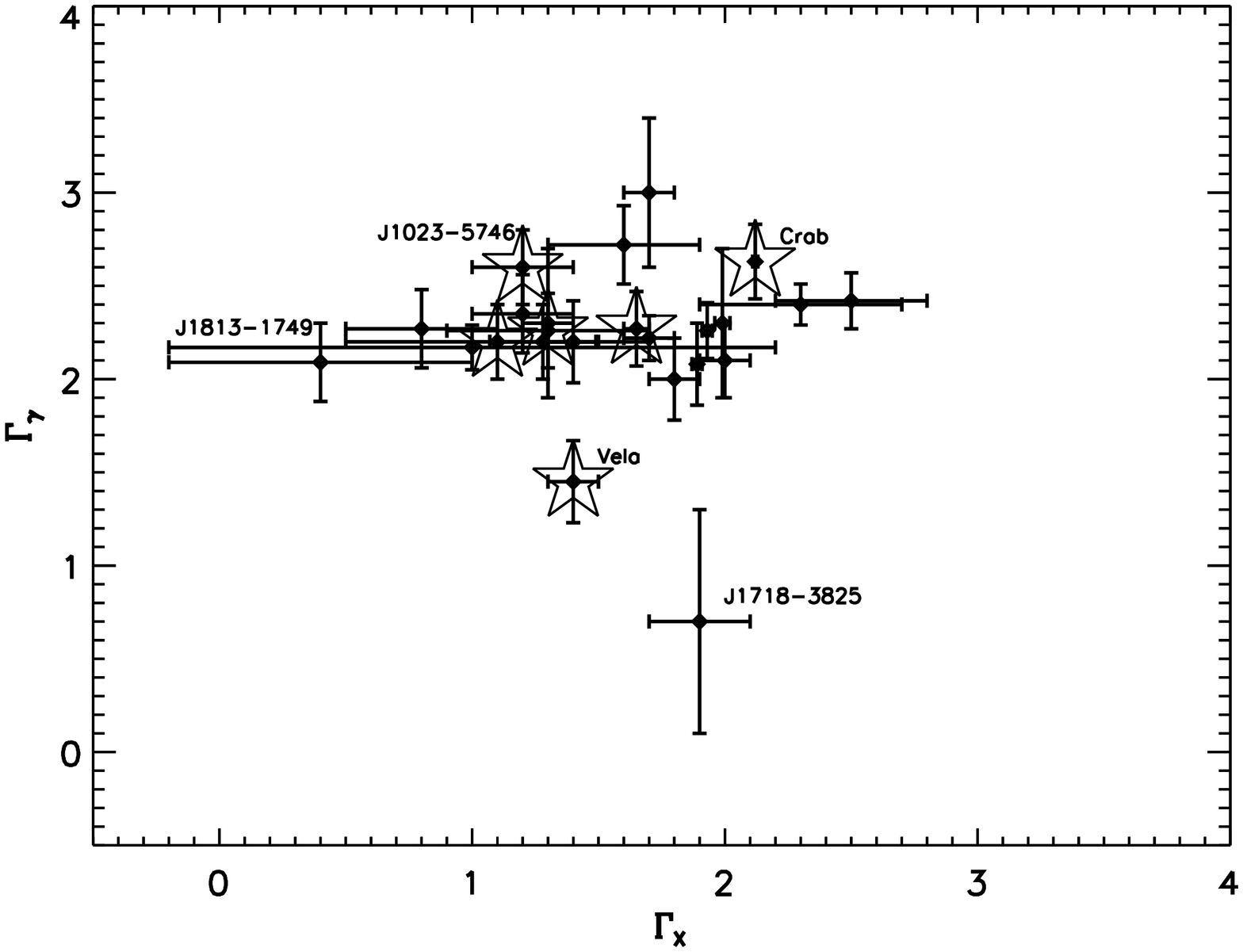}
\includegraphics[width=0.7\textwidth,angle=90]{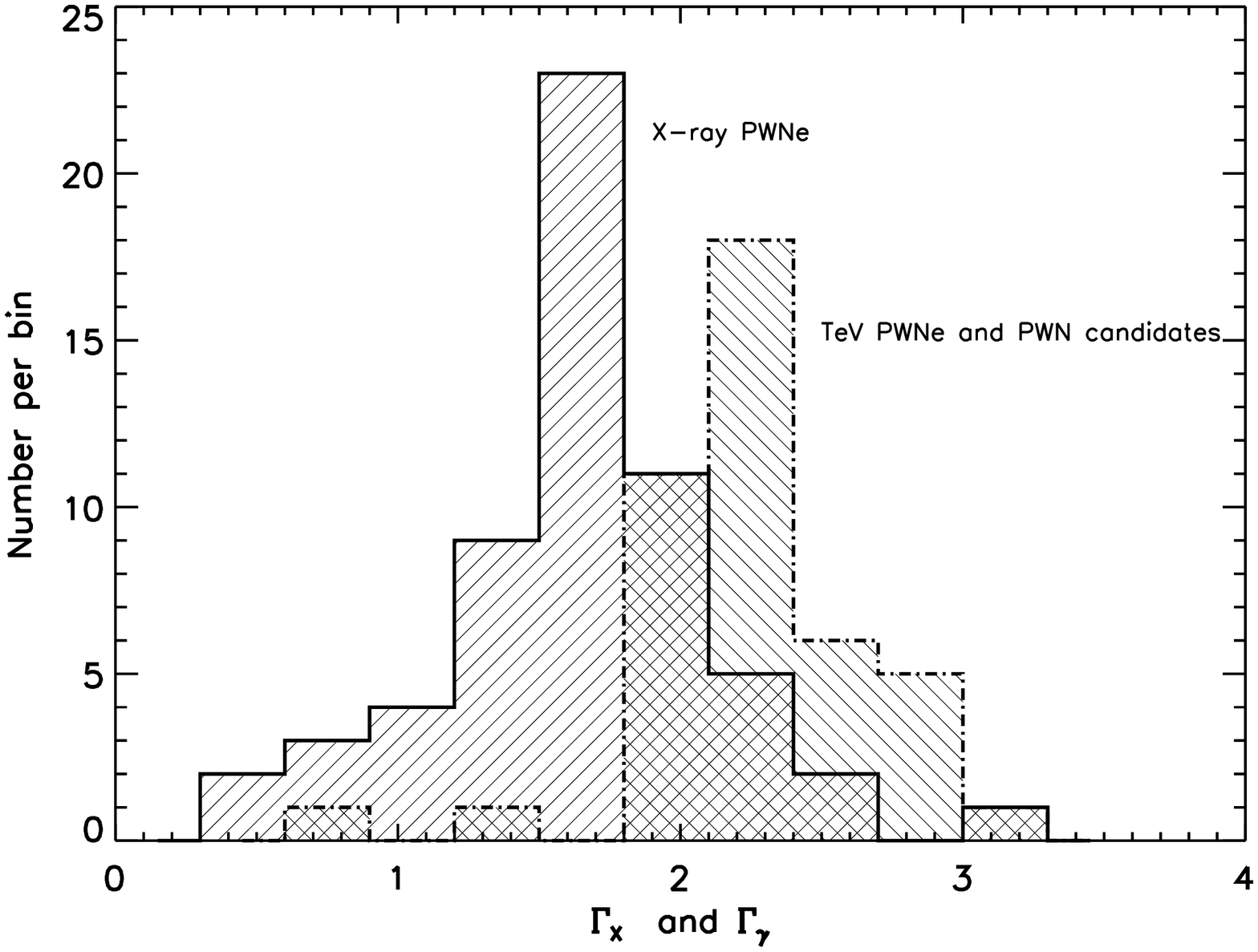}
 \caption{  TeV vs.\ X-ray photon indices for PWNe and PWN candidates   ({\em top})
 and index distributions 
 ({\em bottom}).  
\vspace{-0.8cm} }
\end{figure}

 \begin{figure}
 \centering
 \includegraphics[width=0.7\textwidth,angle=90]{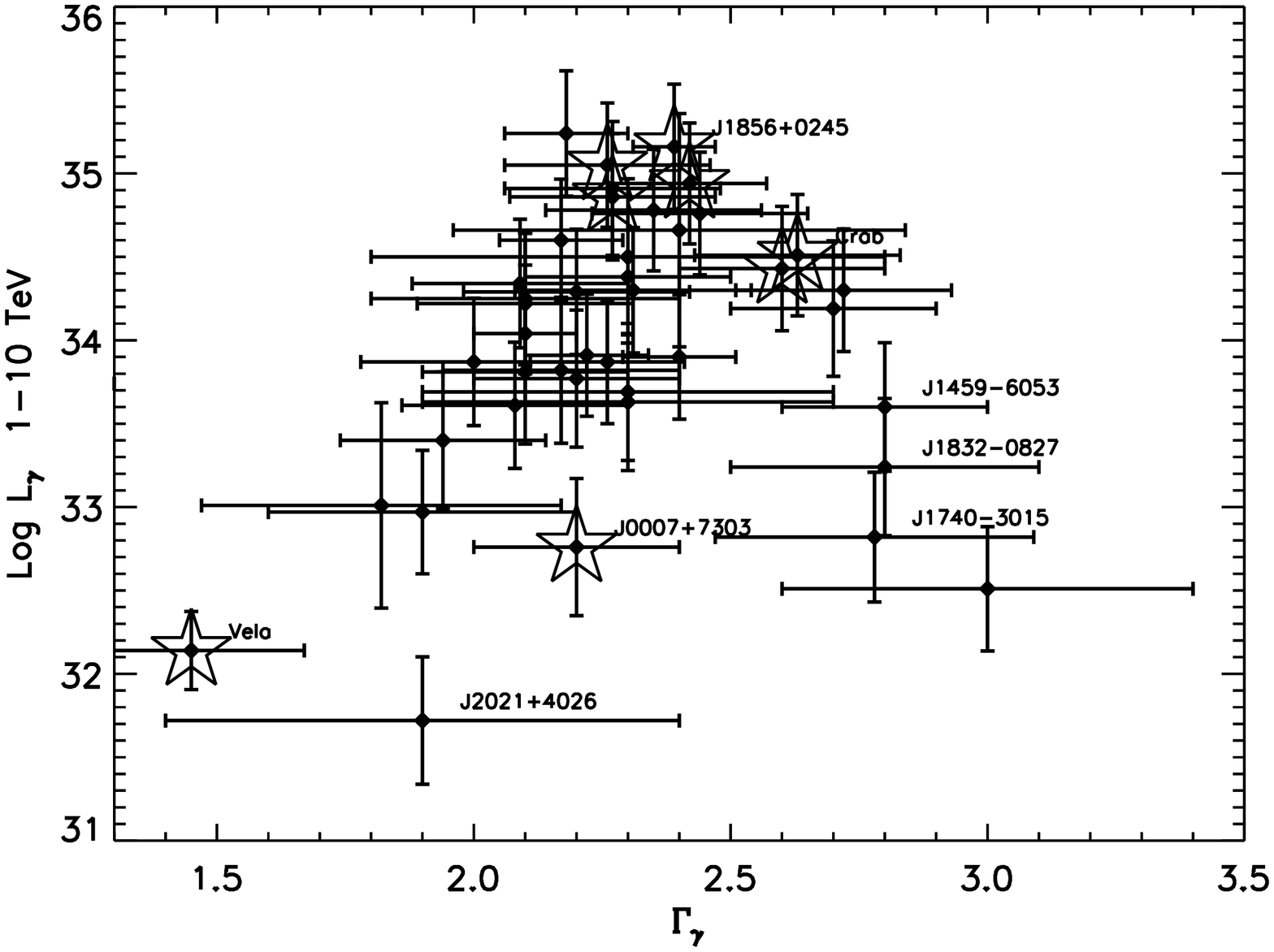}
\includegraphics[width=0.7\textwidth,angle=90]{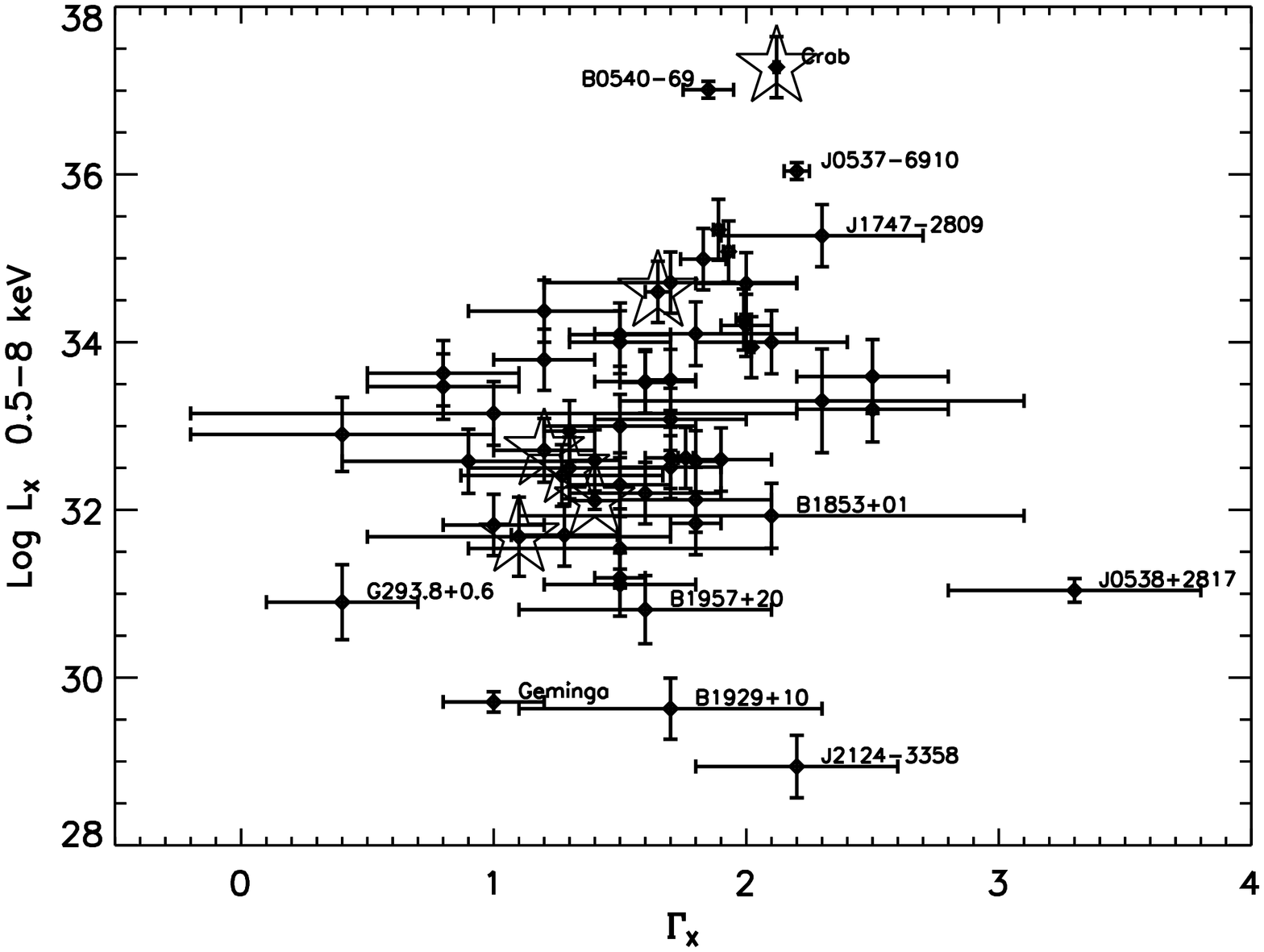}
 \caption{ 
PWN luminosity vs.\ photon index 
 in the TeV  
 and X-ray bands.
 PWNe detected by {\sl Fermi} are marked by stars.
\vspace{-0.8cm} }
\end{figure}

The detected TeV PWNe and PWN candidates can be divided in 
 two classes.
First, relatively small, class includes very
young PWNe powered by energetic pulsars 
(e.g., Crab, J1833--1034, J1930+1852, J1846--0258), without significant offsets
 from the pulsars and X-ray PWNe, in which $\gamma$-rays and X-rays are likely emitted by
the same population of electrons. 
Second class consists of older objects,
in which the TeV PWNe are usually much larger than (and often strongly offset from) the X-ray PWNe
(e.g., HESS J1825--137, HESS J1809--193, Vela-X).   
These relic TeV plerions, offset and compressed by the SNR reverse shock,
 are filled with aged electrons that have cooled due to 
synchrotron and IC energy losses; based on their 
 properties, we can assume
that they still reside inside their host SNRs (which, however, have not been detected for
some of these objects).

TeV emission from PWNe   created by pulsars that
 have escaped from their SNRs and move supersonically in the ISM
have not been firmly detected yet.
 About a dozen of such pulsars are known to be accompanied by ram-pressure confined
bow-shock 
  PWNe with long tails often seen in X-rays and/or radio 
 (e.g., 
 J1747--2958,  
J1509--5850, B0355+54, J0633+1746, B1957+20). 
If the TeV production is due to a hadronic component of the pulsar wind, the nondetections  
might be explained by the
lower ambient density.
 In case of purely leptonic pulsar wind, 
 bow-shock TeV PWNe, which might be created by the freshly shocked electrons in the pulsar
vicinity, are perhaps too faint because of lower spin-down powers of these relatively old
pulsars. 
 A possible explanation of nondections of the long, up to $12'$--$15'$ (or up to $\sim20$  pc), tails filled by the older wind particles  is a lower sensitivity of
 the existing TeV imaging techniques 
 to extended linear structures.
 It is, however, possible that the wind particles channeled into the tails behind the pulsars
 accumulate in lobes, which are not seen
in X-rays (due to the relatively short synchrotron cooling time) but could be
sources of IC TeV emission, strongly offset from the pulsar. 
As the surface brightness   
of such lobes may  be relatively low, deep TeV (and/or radio) observations are required
for their detection.

Multiwavelength observations of PWNe are crucial because they provide identifications for VHE sources and reveal the energetics and composition of pulsar winds.  In addition to  X-ray and TeV observations, the IC PWN component can be detected with {\sl Fermi } LAT in the GeV band, where even old objects should exhibit uncooled IC spectra, matching the radio synchrotron component. The data accumulation must be complemented by development of multi-zone models of PWN evolution \citep{2012ApJ...752...83T}, to understand the nature of pulsar winds and their role in seeding the Galaxy with energetic particles and magnetic fields.

\acknowledgements
This material is  based upon work supported by National Science Foundation under grants AST-0908733 and AST-0908611. The work of GGP was partially supported by the Ministry of Education and Science of the Russian Federation (contract 11.G34.310001).

\bibliographystyle{asp2010}  

\bibliography{kargaltsev_bib}

\end{document}